\documentclass[10pt,a4paper]{article}
\usepackage[T1]{fontenc}
\usepackage[utf8]{inputenc}
\usepackage[margin=25mm]{geometry}
\usepackage{mathptmx}
\usepackage{amsmath,amssymb,graphicx,booktabs}
\usepackage{caption,microtype,url}
\usepackage[hidelinks]{hyperref}
\usepackage{xcolor}
\usepackage{mathtools}

\usepackage[style=numeric, maxcitenames=1, mincitenames=1, maxbibnames=3, backend=biber, sorting=none, giveninits=true]{biblatex}
\renewbibmacro{in:}{}
\DeclareFieldFormat[article]{pages}{#1}

\newcommand{\dcp}{\ensuremath{\delta_{\mathrm{CP}}}}
\newcommand{\echi}{\ensuremath{\Delta\chi^2_{\mathrm{CPV}}}}

\begin{document}
\setlength{\abovedisplayskip}{2pt}
\setlength{\belowdisplayskip}{2pt}

\begin{center}
{\Large\bfseries Machine learning assisted $\nu/\bar{\nu}$ discrimination
in atmospheric neutrino events for CPV sensitivity in a 20-kton liquid scintillator detector\par}
\vspace{0.5em}
Dek Hao Liew$^{1}$, Mohammad Nizam$^{1,2,*}$, Shu Tian Eu$^{1}$, Siew Yan Hoh$^{1}$\par
\vspace{0.3em}
{\small $^{1}$Department of Physics, Xiamen University Malaysia, Sepang, Selangor, Malaysia\par}
{\small $^{2}$Santa Cruz Institute for Particle Physics, University of California Santa Cruz, Santa Cruz, California, USA\par}
{\small $^{*}$Corresponding author: mohammad.nizam@xmu.edu.my\par}
\end{center}

\begin{abstract}
\noindent\textbf{Purpose:} Atmospheric neutrinos span a broad range of energies and baselines, offering a complementary probe of leptonic CP violation. However, a JUNO-like liquid scintillator detector without a magnetic field cannot directly separate neutrino and antineutrino interactions, reducing CP violation sensitivity. This study investigates whether machine learning based statistical separation can improve CP violation sensitivity in a 20-kton detector.

\vspace{0.2em}\noindent\textbf{Methods:} $\nu_{\mu}$ and $\nu_e$ charged current events are generated with GENIE (using Honda flux) in a JUNO-like detector geometry and are reweighted using OscProb three flavor oscillation probabilities incorporating Earth matter effects and NuFIT parameters. Detector response is approximated by smearing lepton and hadronic energies, neutrino direction, and neutron multiplicity. Logistic regression, XGBoost, and multilayer perceptron (MLP) classifiers are trained on inelasticity, neutron multiplicity, and charged lepton energy, with XGBoost selected for the analysis. CP violation sensitivity is evaluated for a 200 kton-year exposure using a Poisson likelihood with systematic nuisance parameters and marginalization over five oscillation parameters within their respective $3\sigma$ ranges.

\vspace{0.2em}\noindent\textbf{Results:} XGBoost achieves an area under the receiver operating characteristic curve (AUC) of approximately 0.83 for both flavor samples. Under the square root approximation, peak CP violation sensitivity improves from 1.49 $\sigma$ to 1.96 $\sigma$ for normal mass ordering and from 1.14 $\sigma$ to 1.61 $\sigma$ for inverted mass ordering.

\vspace{0.2em}\noindent\textbf{Conclusion:} Machine learning based statistical neutrino-antineutrino separation can recover part of the CP violation sensitivity lost in charge blind detectors. These preliminary results motivate further studies incorporating validated detector response and systematic uncertainties.
\end{abstract}

\noindent \textbf{Acknowledgments:} This work was supported by Xiamen University Malaysia (XMUM) under the XMUM Research Fund (XMUMRF) grant number: XMUMRF/2025-C15/IPHY/0007.

\section{Introduction}

Atmospheric neutrinos span broad energy and Earth crossing baselines, providing a complementary probe of leptonic CP violation (CPV) \cite{Birkenfeld2026}. However, unmagnetized liquid scintillator (LS) detectors cannot directly determine the outgoing lepton charge, leaving neutrino and antineutrino interactions statistically mixed within each flavor sample. Since their oscillation probabilities and interaction characteristics differ, this mixing dilutes the observable CP violating effects. Recent studies demonstrate liquid scintillator detectors can discriminate atmospheric neutrinos from antineutrinos using machine learning (ML), where inelasticity, neutron multiplicity, and event topology features provide complementary discrimination power for charged current (CC) interactions \cite{He2026,Liu}. These results motivate investigating whether statistical separation of neutrino and antineutrino improves CPV sensitivity in a generic 20-kton JUNO-like detector using simulated atmospheric $\nu_\mu$ CC and $\nu_e$ CC events. We compare three ML classifiers trained on inelasticity, neutron multiplicity, and charged lepton energy. Ultimately, the XGBoost classifier is selected for the CPV sensitivity analysis after applying detector response smearing, systematic uncertainties, and marginalization over oscillation parameters. For a 200 kton-year exposure of a LS detector, we quantify the sensitivity improvement relative to an unclassified sample for both normal (NO) and inverted (IO) mass orderings, providing a preliminary benchmark for ML assisted atmospheric neutrino CPV searches.

\section{Simulation and analysis methodology }

\subsection{Event generation, oscillations, and detector response} 

Unoscillated atmospheric neutrino CC events with carbon and hydrogen nuclei (88\% and 12\% by mass) as targets are generated with GENIE using the JUNO site Honda flux tables \cite{Honda2015} ($0.1 \leq E_\nu \leq 100$~GeV). Three flavor oscillation probabilities, including Earth matter effects, are calculated with OscProb, using the NuFIT 6.1 best-fit parameters for NO and IO. These probabilities are then used to weight each unoscillated $\nu_\mu$ and $\nu_e$ event, yielding the expected event distributions for each flavor. To approximate detector resolutions, the true charged lepton ($E_\ell$) and hadronic ($E_{\mathrm{had}}$) energies from GENIE are independently smeared using Gaussian resolutions of 5\% and 20\%, respectively, to reconstruct the neutrino energy as $E_{\mathrm{rec}} = E_\ell + E_{\mathrm{had}}$. Events with $0.1 \leq E_{\mathrm{rec}} \leq 10$~GeV are retained for the analysis. True neutrino directions are Gaussian smeared to obtain an energy dependent angular resolution of $5^\circ$--$28^\circ$ \cite{Li_2026}. From the smeared energies, the reconstructed inelasticity is computed as $y = E_{\mathrm{had}} / (E_{\mathrm{had}} + E_\ell)$. The neutron multiplicity, $n_{fn}$, is modeled by applying a 90\% binomial tagging efficiency to the true final state neutron count \cite{Liu}. These simplified parameterizations provide an initial estimate of the detector performance, while full optical simulation and detailed event reconstruction will be required for a realistic assessment.

\subsection{Machine learning classification}

\begin{figure}[htbp]
\centering
\includegraphics[width=0.7\linewidth]{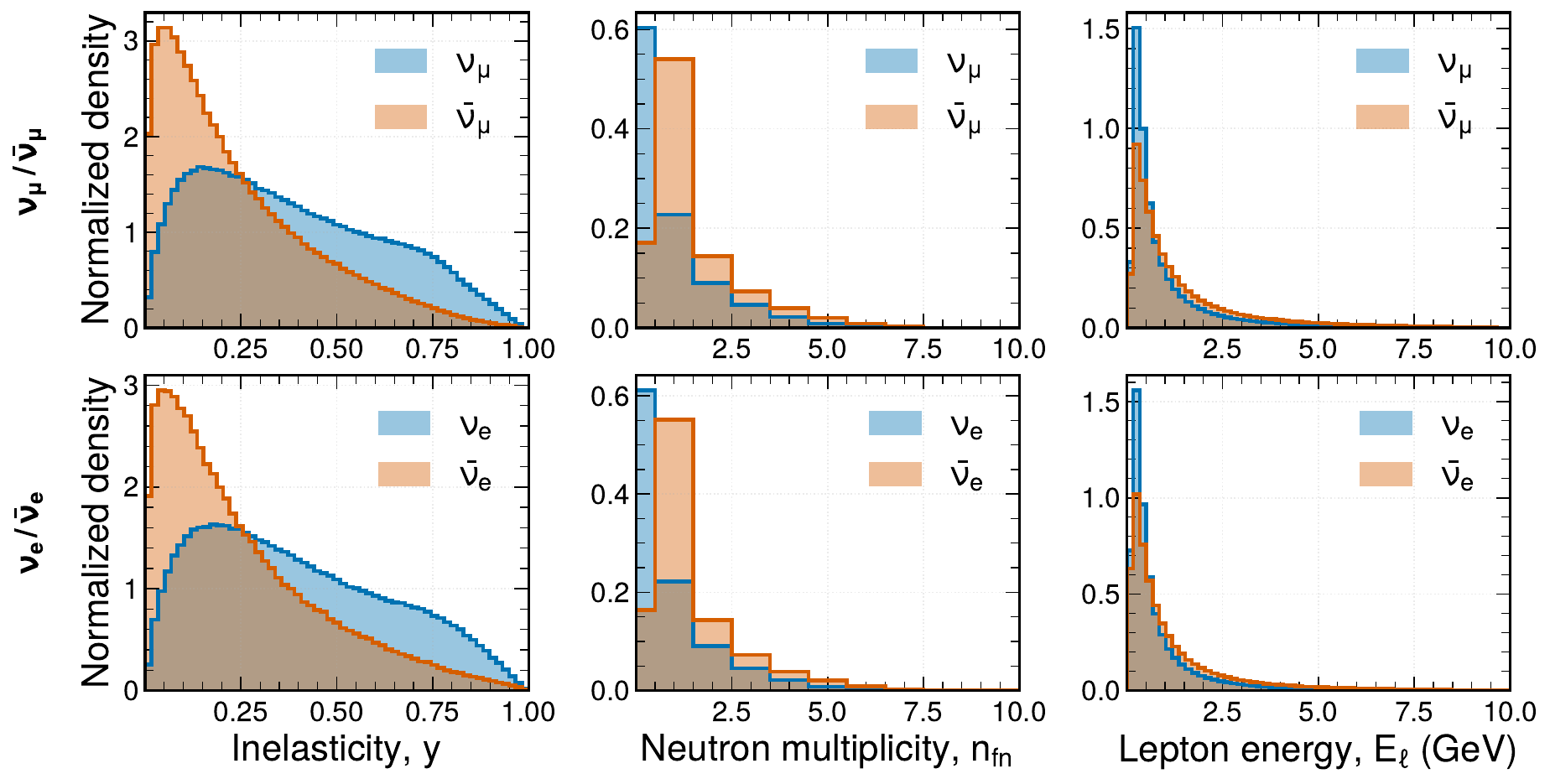}\hfill
\caption{Normalized distributions of input features: inelasticity ($y$), neutron multiplicity ($n_{fn}$), and charged lepton energy ($E_\ell$).}
\label{fig:cpv}
\end{figure}

Logistic regression, XGBoost, and multilayer perceptron (MLP) are trained on the reconstructed features $y$, $n_{fn}$, and $E_\ell$ using independent datasets of approximately $4.0 \times 10^6$ ($\nu_\mu$) and $2.0 \times 10^6$ ($\nu_e$) CC events, partitioned into 60\% for training, 15\% for validation, 10\% for calibration, and 15\% for testing. Logistic regression provides a linear baseline, while XGBoost and the MLP can capture nonlinearities. Figure \ref{fig:cpv} shows the normalized feature distributions for $\nu$ and $\bar{\nu}$. There are differences in their distributions that provide statistical discrimination, although a substantial amount of overlap remains between them.

\begin{figure}[htbp]
\centering
\includegraphics[width=0.6\linewidth]{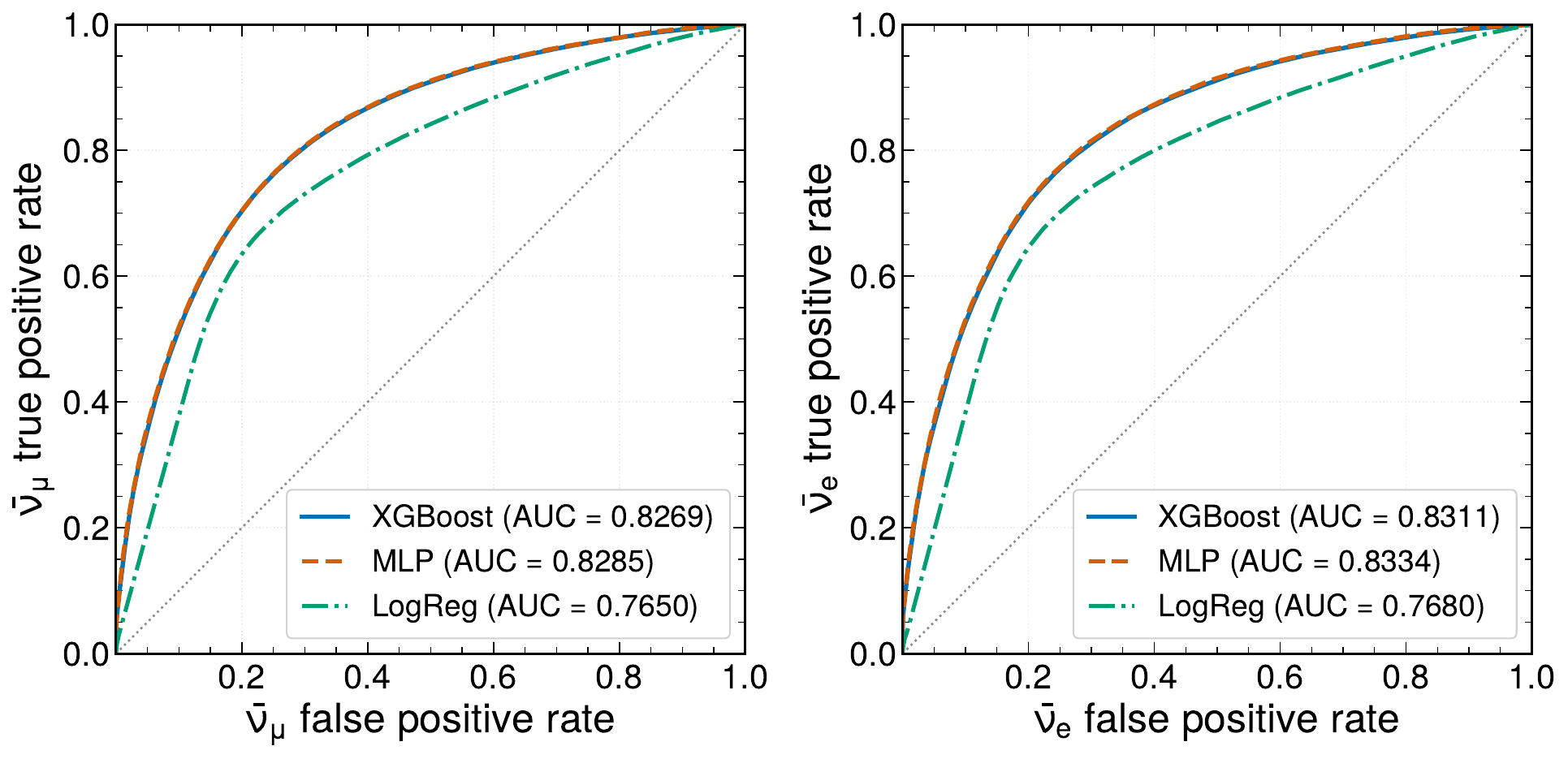}
\caption{ROC curves comparing Logistic Regression, XGBoost, and MLP for the muon (left) and electron (right) flavors.}
\label{fig:roc}
\end{figure}

\noindent The classification performance of the three models is compared using the receiver operating characteristic (ROC) curves shown in Figure \ref{fig:roc}. XGBoost and the MLP clearly outperform the linear logistic regression baseline, demonstrating the importance of capturing nonlinear feature correlations. However, XGBoost is selected for the sensitivity analysis based on its classification performance and computational efficiency.

\noindent Table 1 shows that $n_{fn}$ provides the largest contribution to the XGBoost classification, followed by $y$ and $E_\ell$. The reported percentages represent the relative feature importance within the trained model, not a direct measure of the independent physical information carried by each observable.

\begin{table}[htbp]
\centering
\begin{tabular}{lcc}
\hline
Features & $\nu_\mu/\bar{\nu}_\mu$ (\%) & $\nu_e/\bar{\nu}_e$ (\%) \\ 
\hline
Inelasticity ($y$)              & 17.1 & 15.5 \\
Neutron Multiplicity ($n_{fn}$) & 77.3 & 79.2 \\
Charged Lepton Energy ($E_\ell$)& 5.6  & 5.2  \\
\hline
\end{tabular}
\caption{Relative XGBoost feature importance gain (\%) for $\nu_\mu/\bar{\nu}_\mu$ and $\nu_e/\bar{\nu}_e$ event classification. Values reflect the fractional contribution of each input observable to the decision tree splits.}
\label{tab:feature_importance}
\end{table}

\subsection{CP violation sensitivity}

The sensitivity is evaluated for a 200 kton-year exposure for a 20-kton LS detector under three scenarios: a charge blind sample combining neutrinos and antineutrinos per flavor, an ML assisted analysis sorting events into statistically enriched $\nu$-like and $\bar{\nu}$-like categories via flavor specific calibrated XGBoost score thresholds (including misclassifications), and an idealized scenario with perfect neutrino-antineutrino discrimination. 

To construct the statistical test for these scenarios, events are first binned into 10 non-uniform bins in reconstructed neutrino energy ($E_{\mathrm{rec}}$), over $0.1$--$10\text{ GeV}$, with bin edges defined at $[0.1$, $0.2$, $0.4$, $0.6$, $0.8$, $1.0$, $1.5$, $2.0$, $3.0$, $6.0$, $10.0]\text{ GeV}$, and 10 uniform bins in reconstructed zenith angle ($\cos\theta_z^{\mathrm{rec}} \in [-1, 1]$). An Asimov dataset provides the expected counts at each assumed true value of $\delta_{\mathrm{CP}}$ by scaling the Monte Carlo events to the 20,790 ($\nu_\mu$ CC + $\nu_e$ CC) interactions expected for a 200 kton-year exposure \cite{Guo2017}. The CPV sensitivity is evaluated using a Poisson likelihood ratio ($\chi^2$) statistic, with systematic uncertainties incorporated through nuisance parameters \cite{Choubey2014}:
\begin{equation}
\chi^2(\delta_{\mathrm{CP}},\boldsymbol{\theta},\boldsymbol{\eta}) = 2\sum_i\left[\mu_i-n_i+n_i\ln\left(\frac{n_i}{\mu_i}\right)\right] + \sum_k\left(\frac{\eta_k}{\sigma_k}\right)^2 + \chi^2_{\mathrm{prior}}(\boldsymbol{\theta}), \qquad \mu_i = \mu_i^0(\delta_{\mathrm{CP}},\boldsymbol{\theta})\left(1+\sum_k f_i^k\eta_k\right),
\label{eq:likelihood}
\end{equation}

\noindent where $n_i$ and $\mu_i$ are the Asimov and predicted counts in bin $i$, respectively, $\mu_i^0$ is the prediction with all $\eta_k = 0$, and $\eta_k$ denotes a nuisance parameter with prior width $\sigma_k$. The coefficient $f_i^k$ describes the fractional effect of nuisance parameter $k$ on bin $i$, assuming a linear response and uncorrelated nuisance parameters. The vector $\boldsymbol{\theta}$ denotes the five remaining oscillation parameters, and $\chi^2_{\mathrm{prior}}$ is their NuFIT~6.1 prior term, defined below.

Seven systematic uncertainties are included: flux normalization (20\%), cross section normalization (10\%), detector normalization (5\%), zenith angle tilt (5\%), energy tilt (5\%)~\cite{Choubey2014}, and the $(\nu_e/\bar{\nu}_e)$ (5\%) and $(\nu_\mu/\bar{\nu}_\mu)$ flux ratios (2\% below 1 GeV and 6\% above 1 GeV) \cite{Wendell2010}. The three normalizations scale all bins equally, while the zenith and energy tilts vary as $\cos\theta_z$ and $\ln(E/2~\mathrm{GeV})$, respectively, evaluated at the bin centers~\cite{GonzalezGarcia2004}. Each flux ratio acts only on events from the corresponding $\bar{\nu}$ flux; the $\nu_e/\bar{\nu}_e$ uncertainty is energy independent, whereas the $\bar{\nu}_\mu/\nu_\mu$ uncertainty steps at 1~GeV. The two ratios are treated as uncorrelated. The CP violation statistic tests both CP conserving phases, $\delta_{\rm CP}=0$ and $\pi$, against each assumed true value of $\delta_{\rm CP}$, separately for NO and IO:
\begin{equation}
\echi(\dcp^{\mathrm{true}})=\min_{\dcp^{\mathrm{test}}\in\{0,\pi\}}
\chi^2_{\mathrm{prof}}\big(\dcp^{\mathrm{test}};\,\dcp^{\mathrm{true}}\big).
\end{equation}

Here, $\chi^2_{\mathrm{prof}}$ is the profiled statistic, minimized over $\boldsymbol{\theta}$ and $\boldsymbol{\eta}$. The five remaining oscillation parameters are profiled within their $3\sigma$ intervals from NuFIT~6.1, with asymmetric Gaussian priors from the corresponding $1\sigma$ errors. The $\theta_{23}$ prior is centered at the best fit value and therefore disfavors the opposite octant. The octant degeneracy is therefore not considered in this study. The preliminary approximate significance is estimated as $n_{\sigma}\simeq\sqrt{\scriptstyle{\Delta\chi^2_{\mathrm{CPV}}}}$.

\section{Results and discussion}

Table \ref{tab:cpresults} summarizes the peak CPV sensitivity for a 200 kton-year exposure under three scenarios. The ML assisted method improves the peak CPV sensitivity $\Delta\chi^2_{\mathrm{CPV}}$ from 2.22 ($1.49 \, \sigma$)  to 3.86 ($1.96 \, \sigma$) for NO and from 1.29 ($1.14 \, \sigma$) to 2.58 ($1.61 \, \sigma$) for IO as compared to the charge blind case. The ideal separation scenario provides an upper benchmark with perfect $\nu/\bar{\nu}$ identification producing  peak $\Delta\chi^2_{\mathrm{CPV}}$ values of 6.98 (NO) and 5.43 (IO).


\begin{table}[htbp]
\centering
\begin{tabular}{lccc}
\toprule
Ordering  & Charge blind  & ML assisted  & Ideal separation \\
\midrule
NO & 2.22 (1.49) & 3.86 (1.96) & 6.98 (2.64)\\
IO & 1.29 (1.14) & 2.58 (1.61) & 5.43 (2.33)\\
\bottomrule
\end{tabular}
\caption{Peak CPV sensitivity for an exposure of 200 kton-year. Values in parentheses are the approximate significance.}
\label{tab:cpresults}
\end{table}

\begin{figure}[htbp]
\centering
\includegraphics[width=0.47\linewidth]{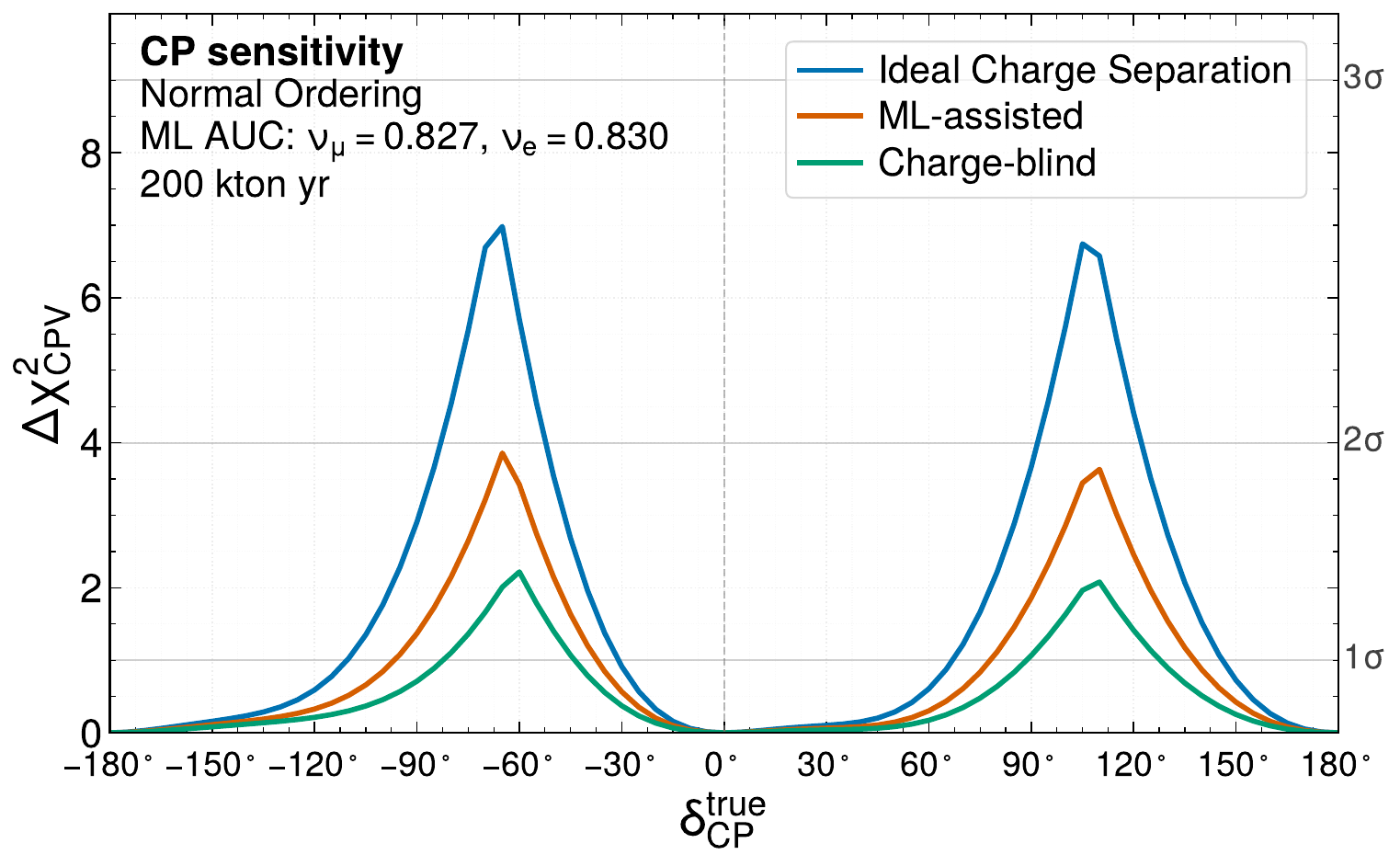}\hspace{0.1cm}
\includegraphics[width=0.47\linewidth]{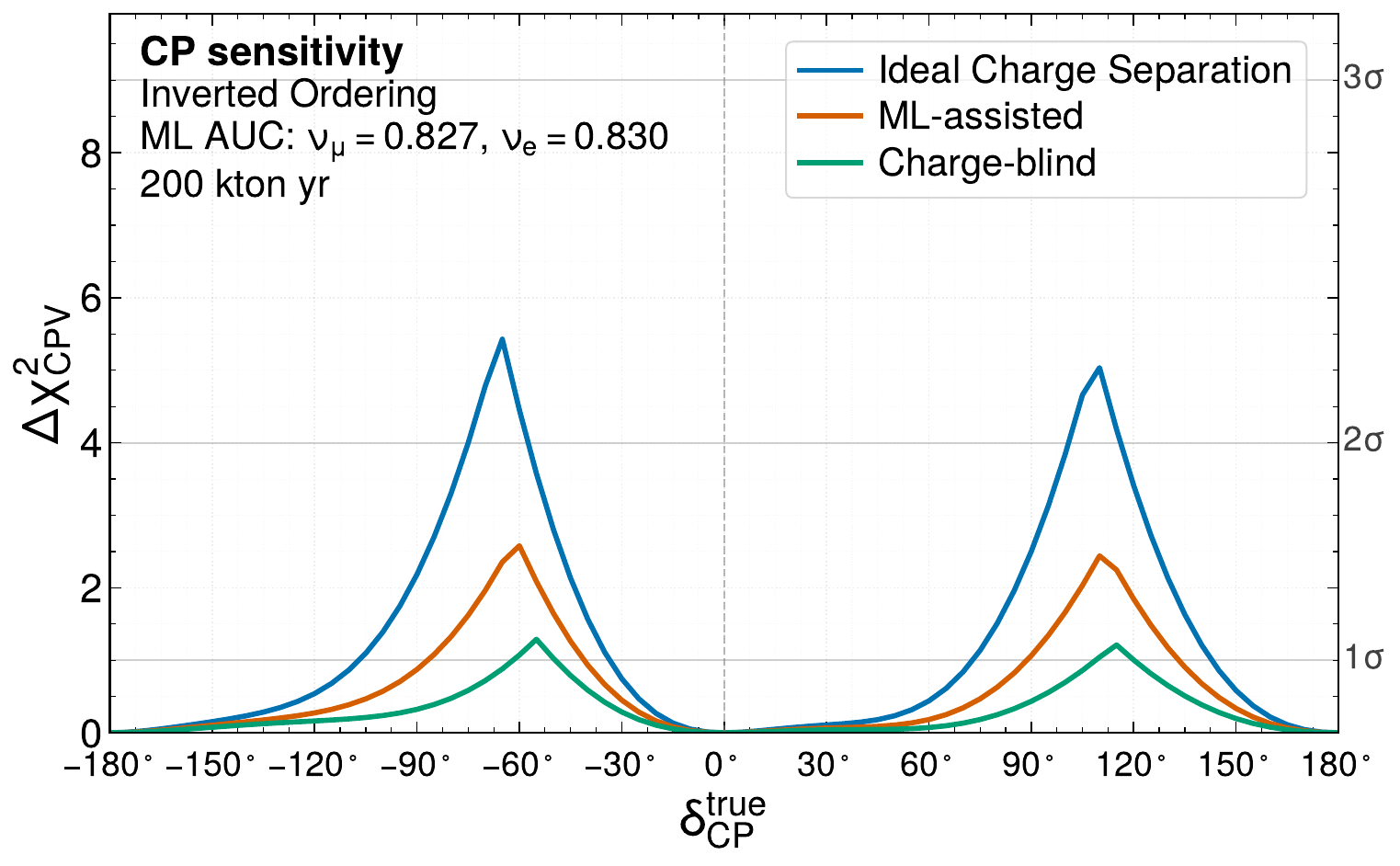}
\caption{Expected CP violation sensitivity $(\Delta\chi^2_{\mathrm{CPV}})$ as a function of the true $CP$ phase $\delta_{\mathrm{CP}}^{\mathrm{true}}$ for a 200 kton-year exposure under normal ordering (left) and inverted ordering (right). In both panels, the ML assisted statistical separation (orange) is compared against the charge blind baseline (green) and the theoretical ideal charge separation (blue).}
\label{fig:cpv1}
\end{figure}

The gap between the ML assisted and ideal separation cases reflects the overlap of the neutrino and antineutrino classifier score distributions. At the adopted classification threshold, the confusion matrix indicates misclassification rates of 19.8\% and 30.0\% for $\nu_\mu$ and $\bar{\nu}_\mu$, and 18.6\% and 30.4\% for $\nu_e$ and $\bar{\nu}_e$, respectively. This cross contamination limits the purity of the enriched samples and reduces the contrast between their oscillation dependent event distributions. Figure \ref{fig:cpv1} shows $\Delta\chi^2_{\mathrm{CPV}}$ as a function of the true $\delta_{\mathrm{CP}}$. The ML assisted analysis enhances sensitivity relative to the charge blind baseline at all CP violating values of $\delta_{\mathrm{CP}}$, partially recovering the lost CP sensitivity, though the magnitude depends on $\delta_{\mathrm{CP}}$ and mass ordering. Profiling the five remaining oscillation parameters reduces the peak $\Delta\chi^2_{\mathrm{CPV}}$ relative to fixed parameters by less than 2\% (NO) and 4--6\% (IO) in all scenarios.

The classifier achieves an AUC of approximately 0.83 for both neutrino flavors.
Neutron multiplicity is the dominant model derived feature (contributing approximately 77\% and 79\% for the muon and electron flavor samples, respectively), while inelasticity and charged lepton energy provide smaller complementary contributions. 

These results are preliminary because the detector response is approximated by parameterized smearing and a simplified neutron tagging model. Therefore, sensitivity estimates depend on the assumed resolutions, tagging efficiency, and systematic uncertainties. More realistic modeling of neutron detection, event reconstruction, backgrounds, and correlated systematics is required to assess the robustness of the improvement. Within these limitations, the study shows that ML based statistical discrimination is a promising way to improve CPV sensitivity of a large LS detector.



\section{Summary and Outlook}

This study investigates the potential of machine learning assisted statistical neutrino-antineutrino discrimination to improve the atmospheric neutrino CPV sensitivity in a generic 20-kton liquid scintillator detector. The XGBoost classifier achieves an AUC of approximately 0.83 for both muon and electron flavor samples, with neutron multiplicity providing the largest contribution. For a 200 kton-year exposure, the peak approximate CPV sensitivity increases from 1.49 $\sigma$ to 1.96 $\sigma$ for NO and from 1.14 $\sigma$ to 1.61 $\sigma$ for IO, relative to the charge blind case. While these preliminary results demonstrate that statistical discrimination can recover part of the diluted CPV sensitivity, future work will incorporate fully validated detector simulations, realistic neutron reconstruction, and additional topological observables to comprehensively assess the experimental feasibility of this approach.


\section*{Declarations}

\noindent\textbf{Conflict of interest:} The authors declare no competing financial or non-financial interests.

\noindent\textbf{Data availability:} The analysis code is available on Zenodo: \href{https://doi.org/10.5281/zenodo.24053212}{doi:10.5281/zenodo.24053212}.

\printbibliography

@article{Guo2017,
  author  = {Guo, W. and others},
 % title   = {Atmospheric neutrinos in {JUNO}},
  journal = {J. Phys. Conf. Ser.},
  volume  = {888},
  pages   = {012205},
  year    = {2017}
}

@article{Birkenfeld2026,
  author  = {Birkenfeld, T. and Stahl, A.},
 % title   = {The sensitivity of liquid scintillator detectors to {CP}-violation with atmospheric neutrinos},
  journal = {J. Instrum.},
  volume  = {21},
  pages   = {P04015},
  year    = {2026}
}

@article{He2026,
  author  = {He, X. and Li, G.-s. and Li, Y.-F. and Luo, W. and Wen, L.-j.},
 % title   = {Atmospheric neutrino charged-current interactions at large liquid-scintillator detectors: {I}. Physics of neutrino-antineutrino discrimination},
  journal = {Chin. Phys. C},
  volume  = {50},
  pages   = {093001},
  year    = {2026}
}

@article{Honda2015,
  author  = {Honda, M. and Athar, M. S. and Kajita, T. and Kasahara, K. and Midorikawa, S.},
%  title   = {Atmospheric neutrino flux calculation using the {NRLMSISE-00} atmospheric model},
  journal = {Phys. Rev. D},
  volume  = {92},
  pages   = {023004},
  year    = {2015}
}

@article{Choubey2014,
  author  = {Choubey, Sandhya and Ohlsson, Tommy},
%  title   = {Bounds on non-standard neutrino interactions using {PINGU}},
  journal = {Phys. Lett. B},
  volume  = {739},
  pages   = {357--364},
  year    = {2014}
}

@article{Wendell2010,
  author  = {Wendell, R. and others},
%  title   = {Atmospheric neutrino oscillation analysis with sub-leading effects in {Super-Kamiokande} {I}, {II}, and {III}},
  journal = {Phys. Rev. D},
  volume  = {81},
  pages   = {092004},
  year    = {2010}
}

@article{GonzalezGarcia2004,
%  title = {Atmospheric neutrino oscillations and new physics},
  author = {Gonzalez-Garcia, M. C. and Maltoni, Michele},
  journal = {Phys. Rev. D},
  volume = {70},
%  issue = {3},
  pages = {033010},
  numpages = {12},
  year = {2004},
%  month = {Aug},
  publisher = {American Physical Society}
%  doi = {10.1103/PhysRevD.70.033010},
%  url = {https://link.aps.org/doi/10.1103/PhysRevD.70.033010}
}

@article{Liu,
author = {Liu, Jiaxi and Zeng, Fanrui and Duyang, Hongyue and Guo, Wanlei and He, Xinhai and Li, Teng and Liu, Zhen and Luo, Wuming and Ma, Wing Yan and Tan, Xiaohan and Wen, Liangjian and Yang, Zekun and Zhang, Yongpeng},
 % title = {Neutrino type identification for atmospheric neutrinos in a large homogeneous liquid scintillation detector},
  
  journal = {Phys. Rev. D},
  volume = {112},
%  issue = {1},
  pages = {012018},
  numpages = {12},
  year = {2025},
%  month = {Jul},
  publisher = {American Physical Society}
%  doi = {10.1103/fznt-z257},
%  url = {https://link.aps.org/doi/10.1103/fznt-z257}
}

@article{Li_2026,
% title = {First attempt of directionality reconstruction for atmospheric neutrinos in a large homogeneous liquid scintillator detector},
  author = {Yang, Zekun and Liu, Jiaxi and Duyang, Hongyue and Guo, Wanlei and He, Xinhai and Hu, Yuxiang and Li, Teng and Liu, Zhen and Luo, Wuming and Luo, Xiaojie and Ma, Wing Yan and Tan, Xiaohan and Zeng, Fanrui and Zhang, Yongpeng},
  journal = {Phys. Rev. D},
  volume = {109},
%  issue = {5},
  pages = {052005},
  numpages = {12},
  year = {2024},
%  month = {Mar},
  publisher = {American Physical Society},
%  doi = {10.1103/PhysRevD.109.052005},
%  url = {https://link.aps.org/doi/10.1103/PhysRevD.109.052005}
}
\end{document}